%% file: main.tex
\documentclass[orivec]{llncs}

\usepackage{wrapfig}
\usepackage{amsmath,amssymb}
\usepackage{listings}
\usepackage{tikz}
\usetikzlibrary{arrows,mmt,backgrounds,fit}
\usetikzlibrary{decorations.pathmorphing}

\usepackage{paralist}

\usepackage[bookmarks,linkcolor=red,citecolor=blue,urlcolor=gray,colorlinks,breaklinks,bookmarksopen,bookmarksnumbered]{hyperref}

\usepackage{basics}
\usepackage{local}

\title{A Universal Machine for Biform Theory Graphs}
\author{Michael Kohlhase \and Felix Mance \and Florian Rabe}

\institute{Computer Science, Jacobs University Bremen\\
\email{initial.lastname@jacobs-university.de} }

\begin{document}
  \maketitle
  \begin{abstract}
    Broadly speaking, there are two kinds of semantics-aware assistant systems for mathematics: proof assistants express the semantic in logic and emphasize
    deduction, and computer algebra systems express the semantics in programming languages and emphasize computation. Combining the complementary strengths of both approaches while mending their
    complementary weaknesses has been an important goal of the mechanized mathematics
    community for some time.

    We pick up on the idea of biform theories and interpret it in the \mmt/ \omdoc
    framework which introduced the foundations-as-theories approach, and can thus
    represent both logics and programming languages as theories. This yields a formal,
    modular framework of biform theory graphs which mixes specifications and
    implementations sharing the module system and typing information.
    
    We present automated
    knowledge management work flows that interface to existing specification/programming
    tools and enable an \openmath Machine, that operationalizes biform theories,
    evaluating expressions by exhaustively applying the implementations of the respective
    operators. We evaluate the new biform framework by adding implementations to the
    \openmath standard content dictionaries.
  \end{abstract}

\section{Introduction}
  \input{intro}

\section{Representing Languages in MMT}
In this section we introduce the \mmt language and directly apply it to modeling the
pieces of our running example by representing \openmath and Scala in \mmt.

  \subsection{The MMT Language and System}
    \input{mmt}
  \subsection{Content Dictionaries as MMT Theories}\label{sec:mmt-om}
    \input{openmath}
    \input{mmt-openmath}
  \subsection{Scala Classes as MMT Theories}
    \input{scala}
    \input{mmt-scala}

\section{Biform Theory Development in MMT}
   \input{mmt-biform}

\section{Mechanizing Biform Theory Graphs}
   \input{rules}

\section{Building a Biform Library}\label{sec:library}
   \input{library}

\section{Conclusion}
  \input{conc}

\paragraph{Acknowledgements}
The work reported here was prompted by discussions with William Farmer and Jacques Carette.
An initial version of the universal machine was developed in collaboration with Vladimir Zamdzhiev.

\bibliographystyle{alpha}
\input{main-final.bbl}

\end{document}

%% file: intro.tex

It is well-known that mathematical practices -- conjecturing, formalization, proving,
etc. -- combine (among others) axiomatic reasoning with computation.  Nevertheless,
assistant systems for the semantics-aware automation of mathematics can be roughly divided
into two groups: those that use \emph{logical languages} to express the semantics and
focus on \emph{deduction} (commonly called \defemph{proof assistants}), and those that use
\emph{programming languages} to express the semantics and focus on \emph{computation}
(commonly called \defemph{computer algebra systems}).
Combining their strengths is an important objective in mechanized mathematics.

Our work is motivated by two central observations.
Firstly, combination approaches often take a deduction or computation system and try to embed the respective other mode into its operations, e.g., \cite{harrison-cas,Maple-Mode} and \cite{kenzo_integrating}, respectively.
Secondly, most of these systems are usually based on the \emph{homogeneous} method, which fixes one foundation
(computational or deductive) with all primitive notions (e.g., types, axioms, or
programming primitives) and uses only conservative extensions (e.g., definitions,
theorems, or procedures) to model domain objects.

In this paper, we want to employ the \emph{heterogeneous} method, which
focuses on encapsulating primitive notions in theories and considers truth
relative to a theory.  It optimizes reusability by stating every
result in the weakest possible theory and using \emph{theory morphisms} to move results
between theories in a truth-preserving way.  This is often called the \emph{little
  theories} approach~\cite{littletheories}.  In computational systems this is mirrored by using programming languages that relegate much of the functionality to an extensible library infrastructure.

\begin{wrapfigure}r{4cm}\vspace*{-2em}
\begin{tikzpicture}[xscale=1.5]
  \node[thy] (sl) at (0,2) {\cn{SL}};
  \node[thy] (pl) at (2,2) {\cn{PL}};
  \node[thy] (spec) at (0,0) {\cn{Spec}};
  \node[thy] (imp) at (2,0) {\cn{Impl}};
  \node[thy] (simp) at (1,-1) {\cn{Obl}};
  \node[thy] (pf) at (1,1) {\cn{Ref}};
  \draw[meta] (sl) -- (spec);
  \draw[meta] (pl) -- (imp);
  \draw[include] (spec) -- (simp);
  \draw[include] (spec) -- (pf);
  \draw[decorate,decoration={snake,amplitude=2pt},|->]  (imp) -- (simp);
  \draw[decorate,decoration={snake,amplitude=2pt},|->]  (pf) -- (imp);
  \begin{pgfonlayer}{background}
  \node[fill=lightgray,fit={(-.8,-1.35) (1.8,1.35)},rounded corners] {};
\end{pgfonlayer}
\end{tikzpicture}\vspace*{-1.5em}
\end{wrapfigure}

In homogeneous approaches, we usually fix a specification language $\cn{SL}$ and a programming language $\cn{PL}$ and one implementation for each.
In program synthesis, a specification \cn{Spec} is extended (hooked arrows) to a refined specification \cn{Ref}, from which a program can be extracted (snaked arrow).
Both can be visualized by the diagram on the right where dotted arrows denote the written-in relation.
In both cases, the proofs are carried out in a theory of \cn{SL}, and a non-trivial generation step crosses the border between the \cn{SL}-based deduction system (the gray area) and the \cn{PL}-based computation system, e.g., \cite{isabellehol_codegen} generates programs from Isabelle/HOL proofs.

Dually, we find approaches that emphasize \cn{PL} over \cn{SL}.
SML-style module systems and object-orientation can be seen as languages that transform parts of \cn{SL} (namely the type system but not the entailment system) into \cn{PL}.
An example is the transformation of \cn{SL}=UML diagrams into \cn{PL}=Java stubs, which are then refined to a Java program.
Advanced approaches can transform the whole specification into \cn{PL} by enriching the programming language as in \cite{extendedml} or the programming environment as in \cite{key}.

A third approach is to develop a language $\cn{SPL}$ that is strong enough to combine the features of specification and programming.
Several implementations of $\lambda$ calculi have been extended with features of programming languages, e.g., Coq \cite{coqmanual} and Isabelle/HOL \cite{isabellehol}.
The FoCaLiZe language \cite{focalize} systematically combines a functional and object-oriented programming language with a logic, and a compilation process separates the two aspects by producing OCaml and Coq files, respectively.
The source files may also contain raw Coq snippets that are not verified by FoCaLiZe but passed on to Coq.
In Dedukti \cite{dedukti}, rewriting rules are used to enhance a specification language with computational behavior, and the computational aspect can be compiled into a Lua program.

We want to create a heterogeneous framework in which we can represent such homogeneous approaches.
We use the \mmt language \cite{RK:mmt:10}, which extends the heterogeneous method with language-independence inspired by logical frameworks.
The key advantage is that this permit flexibly combining arbitrary specification and programming languages.
In \mmt, we represent both \cn{PL} and \cn{SL} as \mmt theories $\ov{\cn{SL}}$ and $\ov{\cn{PL}}$ (see diagram below), which declare the primitive concepts of the respective language.
The dotted lines are represented explicitly using the \emph{meta-theory} relation, and relatively simple mappings (dashed snaked lines) transform between specifications $\cn{Spec}$ and implementations $\cn{Impl}$ and the corresponding \mmt theories $\ov{\cn{Spec}}$ and $\ov{\cn{Impl}}$.

Typically $\cn{SL}$ and $\cn{PL}$ share some language features, e.g., the type system, which
$\cn{SL}$ enriches with deductive primitives and $\cn{PL}$ with computational primitives.
\mmt can represent this by giving a (possibly partial) morphism $\cn{bifound}:\cn{\ov{SL}}\to\ov{\cn{PL}}$ that embeds $\cn{SL}$ features into $\cn{PL}$.
Via $\cn{bifound}$, $\ov{\cn{Impl}}$ can access both $\ov{\cn{SL}}$ and $\ov{\cn{PL}}$ features, and the fact that $\ov{\cn{Impl}}$ implements $\ov{\cn{Spec}}$ is represented as an \mmt theory morphism (dashed line).

\begin{wrapfigure}{r}{3cm}
\vspace{-3em}
\begin{tikzpicture}[xscale=1.7,yscale=1.5,trans/.style={dashed,decorate,decoration={snake,amplitude=2pt},<->}]
  \node[thy] (sl) at (0,3) {\cn{SL}};
  \node[thy] (pl) at (1,3) {\cn{PL}};
  \node[thy] (spec) at (0,0) {\cn{Spec}};
  \node[thy] (imp) at (1,0) {\cn{Impl}};

  \node[thy] (s) at (0,2) {$\ov{\cn{SL}}$};
  \node[thy] (p) at (1,2) {$\ov{\cn{PL}}$};
  \node[thy] (mspec) at (0,1) {\cn{\overline{Spec}}};
  \node[thy] (mimp) at (1,1) {\cn{\overline{Impl}}};
  
  \draw[meta] (s) -- (mspec);
  \draw[meta] (p) -- (mimp);
  \draw[trans]  (s) -- (sl);
  \draw[trans]  (p) -- (pl);
  \draw[trans]  (mimp) -- (imp);
  \draw[trans]  (mspec) -- (spec);
  \draw[view] (s) -- node[above]{bifound} (p);
  \draw[view] (mspec) -- (mimp);
  \begin{pgfonlayer}{background}
  \node[fill=lightgray,fit={(-0.6,0.4) (1.6,2.6)},inner sep=10pt,rounded corners] (foo) {};
\end{pgfonlayer}
\end{tikzpicture}\vspace*{-1em}
\end{wrapfigure}

Our framework is inspired by the \emph{biform theories} of \cite{Farmer:btc07}, which extend axiomatic theories
with \emph{transformers}: named algorithms that implement axiomatically specified function symbols.
We follow the intuition of heterogeneous biform theories but interpret them in \mmt.
Most importantly, this permits $\cn{SL}$ and $\cn{PL}$ to be arbitrary languages represented in \mmt.

We leverage this by using types and examples in $\ov{\cn{Spec}}$ to generate method stubs and test cases in $\ov{\cn{Impl}}$.
Our interest is not (yet) the corresponding treatment of axioms, which would add formal deduction about the correctness of programs.
In particular, we do not provide a formal definition of the meaning of the computational knowledge other than linking symbols to algorithms via theory morphisms.

As a computational backend, we develop what we call the \emph{universal machine}. It
extends \mmt with a component that collects the individual implementation snippets
occurring in a biform \mmt theory graph and integrates them into a rule-based
rewriting engine.
The universal machine keeps track of these and performs computations by applying the available rules.

In the past, a major practical limitation of frameworks like ours has been the development of large libraries of biform theories.
Here a central contribution of our work is that the \mmt API \cite{rabe:mmtabs:13} (providing, e.g., notations, module system, and build processes) makes it easy to write biform theories in practice.
Moreover, the API is designed to make integration with other applications easy so that the universal machine can be easily reused by other systems.

We evaluate this infrastructure in an extensive case study: We represent a collection of \openmath content dictionaries in \mmt (i.e., $\cn{SL}=\cn{OpenMath}$) and provide implementations for the symbols declared in them using the programming language Scala (i.e., $\cn{PL}=\cn{Scala})$.
The resulting biform theory graph integrates \openmath CDs with the Scala code snippets implementing the symbols.



%% file: mmt.tex
{\mmt} \cite{RK:mmt:10} is a knowledge representation format focusing on modularity
and logic-independence.  It is accompanied by the {\mmt} API, which implements the
language focusing on scalable knowledge management and system interoperability. For our
purposes, the simplified fragment of \mmt (which in particular omits named imports and sharing between imports) given in Figure~\ref{fig:mmt} suffices.

\newcommand{\keyw}[1]{\textsf{#1}}
\newcommand{\nonterminal}[1]{\texttt{#1}}
\newcommand{\name}{\nonterminal{ModuleName}}
\newcommand{\module}{\nonterminal{Module}}
\newcommand{\statement}{\nonterminal{Statement}}
\newcommand{\object}{\nonterminal{Term}}
\newcommand{\notation}{\nonterminal{Notation}}

\begin{tabularfigure}{lcl}{A Fragment of the \mmt Grammar}{fig:mmt}
$\module$ &    $::=$ & $\keyw{theory}\;T\;:\name\;\statement^*$\\
               & $|$ & $\keyw{view}\;V\;:\;\name\to\name\; \statement^*$ \\
$\statement$ & $::=$ & $\keyw{constant}\;c\;[:\object]\;[=\object]\;[\#\notation]$\\
               & $|$ & $\keyw{include}\;\name$\\
$\object$    & $::=$ & $c \;|\; x \;|\; \nonterminal{number} \;|\; \oma{\object^+} \;|\; \ombind{\object}{x^+}{\object}$ \\
$\notation$  & $::=$ & $(\nonterminal{number}[\nonterminal{string}...] \;|\; \nonterminal{string})^*$ \\
\end{tabularfigure}

We will briefly explain the intuitions behind the concepts and then exemplify them in the later sections, where we represent \openmath CDs and Scala classes as \mmt theories.

An \mmt \emph{theory} $\keyw{theory}\; T:M\; \Sigma$ defines a theory $T$ with meta-theory
$M$ consisting of the statements in $\Sigma$.  The \emph{meta-theory} relation between theories is crucial to
obtain logic-independence: The meta-theory gives the language, in which the theory is
written.  For example, the meta-theory of a specification is the specification logic, and
the meta-theory of a program is the programming language -- and the logic and the
programming language are represented as \mmt theories themselves (possibly with further
meta-theories).  Thus, \mmt achieves a uniform representation of logics and programming
languages as well as their theories and programs.

\mmt theories form a category, and an \mmt $\keyw{view}\;V:T_1\to T_2\;
\Sigma$ defines a theory morphism $V$ from $T_1$ to $T_2$ consisting of the statements in
$\Sigma$.  In such a view, $\Sigma$ may use the constants declared in $T_2$ and must
declare one definition for every definition-less constant declared in $T_1$.  Views
uniformly capture the relations ``$T_2$ interprets/implements/models $T_1$''.  For example, if
$T_1$ represents a specification and $T_2$ a programming language, then views $T_1\to T_2$
represent implementations of $T_1$ in terms of $T_2$ (via the definitions in $\Sigma$).

Theories and views are subject to the \mmt module system.  Here we will restrict attention
to the simplest possible case of unnamed inclusions between modules: If a module $T$ contains a statement $\keyw{include}\;S$, then all
declarations of $S$ are included into $T$.

Within modules, \mmt uses constants to represent atomic named declarations.  A constant's optional type and definiens are arbitrary terms.
Due to the freedom of using
special constants declared in the meta-theory, a type and a definiens are sufficient to
uniformly represent diverse statements of formal languages such as function symbols,
examples, axioms, inference rules. Moreover, constants have an optional notation which is
used by \mmt to parse and render objects. We will not go into details and instead explain
notations by example, when we use them later on.

\mmt terms are essentially the \openmath objects~\cite{openmath} formed from the
constants included into the theory under consideration.
This is expressive enough to subsume the abstract syntax of a wide variety of formal systems.
We will only consider the fragment of \mmt terms formed from constants $c$, variables
$x$, numbers literals, applications $\oma{f,t_1,\ldots,t_n}$ of $f$ to the $t_i$, and
bindings $\ombind{b}{x_1,\ldots,x_n}{t}$ where a binder $b$ binds the variables $x_i$ in
the scope $t$.



%% file: openmath.tex
\openmath declares symbols in named content dictionaries that have global scope (unlike
\mmt theories where symbols must be imported explicitly). Consequently, references to symbols must reference the CD and
the symbol name within that CD. The official \openmath CDs~\cite{omcds} are a
collection of content dictionaries for basic mathematics.  For example, the content
dictionary \cn{arith1} declares among others the symbols \cn{plus}, \cn{minus},
\cn{times}, and \cn{divide} for arithmetic in any mathematical structure -- e.g., a
commutative group or a field -- that supports it.

Each symbol has a type using the STS type system~\cite{Davenport:asots00}.  The types
describe what kinds of application (rarely: binding) objects can be formed using the
symbol.  For example, its type licenses the application of \cn{plus} to any sequence of
arguments, which should come from a commutative semigroup.  Moreover, each symbol comes
with a textual description of the meaning of the thus-constructed application, and
sometimes axioms about it, e.g., commutativity in the case of \cn{plus}.



%% file: mmt-openmath.tex
\begin{wrapfigure}r{5.8cm}\vspace*{-2.5em}
\begin{tabular}{|l||l|}\hline
\openmath & \mmt \\\hline\hline
CD                 & theory \\
symbol             & constant\\
property $F$       & constant $\oma{\cn{FMP},F}$\\\hline
\end{tabular}\vspace*{-1.5em}
\end{wrapfigure}
We represent every \openmath CD as an \mmt theory, whose meta-theory is a special \mmt
theory \cn{OpenMath}.  Moreover, every \openmath symbol is represented as an \mmt
constant.  All constants are definition-less, and it remains to describe their types and
notations.  Mathematical properties that are given as formulas are also represented as
\mmt constants using a special type.

\paragraph{Meta-Theory and Type System}
\cn{OpenMath} must declare all those symbols that are used to form the types of \openmath symbols.
This amounts to a formalization of the STS type system~\cite{Davenport:asots00} employed in the \openmath CDs.
However, because the details STS are not obvious and not fully specified, 
we identify the strongest type system that we know
how to formalize and of which STS is a weakening.  Here strong/weak means that the typing
relation holds rarely/often, i.e., every STS typing relation also holds in our weakened
version.
The types in this system are: \begin{inparaenum}[\it i)]
\item \cn{Object}
\item $\oma{\cn{mapsto},\cn{Object},\ldots,\cn{Object},A,\cn{Object}}$ where $A$ is either
\cn{Object} or \cn{naryObject}
\item \cn{binder}.
\end{inparaenum}
Here \cn{binder} is the type of symbols that take a context $C$ and an \cn{Object} in context $C$ and return an \cn{Object}.
This type system ends up being relatively simple and is essentially an
arity-system.\footnote{In fact, we are skeptical whether any fully formal type system for all of \openmath can be more than an arity system.}

\begin{wrapfigure}r{4.8cm}
\vspace{-1em}
\begin{tabular}{|c|}
\hline
\begin{lstlisting}
theory OpenMath
  constant mapsto # 1$\times$... $\to$ 2
  constant Object 
  constant naryObject
  constant binder
  constant FMP
\end{lstlisting}
\\
\hline
\end{tabular}
\caption{\mmt Theory \cn{OpenMath}}\label{fig:openmath-mmt}
\vspace*{-2em}
\end{wrapfigure}

Moreover, we add a special symbol \cn{FMP} to represent mathematical properties as
follows: A property asserting $F$ is represented as a constant with definiens
$\oma{\cn{FMP},F}$.\footnote{We do not use a propositions-as-types representation here
  because it would make it harder to translate \cn{OpenMath} to other languages.}
Intuitively, we can think of \cn{FMP} as a partial function that can only be applied to true formulas.
We do not need symbols for the formation of formulas $F$ because they are treated as normal symbols that are introduced in CDs such as \cn{logic1}.

This results in the following \mmt theory \cn{OpenMath} in
Figure~\ref{fig:openmath-mmt}. There, the notation of \cn{mapsto} means that it takes
first a sequence or arguments with separator $\times$ followed by the separator $\to$ and
one more argument.

\begin{tabularfigure}{|l|l|}{\openmath CDs in \mmt}{fig:omex}\hline
\begin{lstlisting}
theory arith1 : OpenMath
  plus  : naryObject $\to$ Object
      # 1$+$...
  minus : Object $\times$ Object $\to$ Object
      # 1 $-$ 2
  plus  : naryObject $\to$ Object
      # 1$*$...
  $\ldots$
\end{lstlisting}
 &
\begin{lstlisting}
theory NumbersTest : OpenMath
  include arith1
  include fns1
  include set1
  include relations1
  maptest = FMP
    {0,1,2} map (x $\mapsto$ -x*x+2*x+3) = {3,4}
\end{lstlisting}\\\hline
\end{tabularfigure}

\paragraph{Notations}
In order to write \openmath objects conveniently -- in particular, to write the examples mentioned below -- we add notations to all \openmath symbols.
\openmath does not explicitly specify notations for the symbols in the official CDs.
However, we can gather many implied notations from the stylesheets provide to generate
presentation \mathml.  Most of these can be mapped to \mmt notations in a straightforward fashion.
As \mmt notations are one-dimensional, we make reasonable adjustments to two-dimensional
\mathml notations such as those for matrices and fractions.

\begin{example}
We will use a small fragment of our case study (see Section~\ref{sec:library}) as a running example.
The left listing in Fig.~\ref{fig:omex} gives a fragment of the \mmt theory representing the CD \cn{arith1}.
Here the notation of \cn{plus} means that it takes a sequence or arguments with separator
$+$, and the one of \cn{minus} that it takes two arguments separated by $-$.

The right listing uses the module system to import some CDs and then give an example of a true computation as an axiom.
It uses the symbols \cn{set1?set}, \cn{fns1?lambda}, and \cn{relation1?eq} and the notations we declare for them.
\end{example}



%% file: scala.tex
Scala~\cite{scala} combines features of object-oriented and functional programming languages.
At the module and statement level, it follows the object-oriented paradigm and is similar to Java.
At the expression level, it supplements Java-style imperative features with simple function types and inductive types.

A \emph{class} is given by its list of member declarations, and we will only make use of 3
kinds of members: \emph{types}, immutable typed \emph{values}, and \emph{methods}, which
are essentially values of functional type.

Values have an optional definiens, and a class is \emph{concrete} if all members have one,
otherwise abstract.  Scala introduces special concepts that can be used instead of classes
without constructor arguments: \emph{trait} in the abstract and \emph{object} in the
concrete case.  Traits permit \emph{multiple inheritance}, i.e., every class can inherit
from multiple traits.  Objects are singleton classes, i.e., they are at the same time a
class and the only instance of this class. An object and a trait may have the same name,
in which case their members correspond to the static and the non-static members,
respectively, of a single Java class.



%% file: mmt-scala.tex
The representation of Scala classes proceeds very similarly to that of \openmath CDs above
(see Figure~\ref{fig:mmt-scala}).  In particular, we use a special meta-theory \cn{Scala}
that declares the primitive concepts needed for our Scala expressions.  Then we represent
Scala classes as {\mmt} theories and members as constants.
While \openmath CDs always have the flavor of specifications, Scala classes can have the
flavor of specifications (abstract classes/traits) or implementations (concrete classes/objects).

\begin{figure}[ht]\centering
\begin{tabular}{|l||l|}\hline
Scala & \mmt \\\hline\hline
trait $T$          & theory $\enc{T}$ \\
type member        & constant of type \cn{type}\\
value member       & constant \\
method member      & constant of functional type \\\hline
object $O$ of type $T$ & theory morphism $\enc{T}\arr\cn{Scala}$ \\
members of $O$     & assignment to the corresponding $\enc{T}$-constant\\\hline
extension between classes & inclusion between theories \\\hline
\end{tabular}
\caption{Scala Classes as \mmt Theories}\label{fig:mmt-scala}
\end{figure}

\paragraph{Meta-Theory and Type System}
Our meta-theory \cn{Scala} could declare symbols for every primitive concept used in Scala
expressions.
However, most of the complexity of Scala expressions stems from the richness of the term language.
While the representation of terms would be very useful for verification systems, it does not contribute much to our goals of computation and biform development.
Therefore, we focus on the simpler type language.
Moreover, we omit many theoretically important technicalities (e.g., singleton and existential types) that have little practical bearing.
Indeed, many practically relevant types (e.g., function and collection types) are derived notions defined in the Scala library.


\newcommand{\vc}[1]{\mathcal{V}_{#1}}
\newcommand{\scc}[1]{\mathcal{S}_{#1}}

Therefore, we represent only the relevant fragment of Scala in \cn{Scala}.
Adding further features later is easy using the \mmt module system.
For all inessential (sub-)expressions, we simply make use \mmt escaping: \mmt
expressions can seamlessly escape into arbitrary non-\mmt formats.

\begin{wrapfigure}r{4.8cm}
\vspace*{-1.5em}
\begin{lstlisting}[frame=trbl]
theory Scala
  constant type
  constant Any
  constant Function # (1,...)=> 2
  constant Lambda   # (1,...)=> 2
  constant List     # List[1]
  constant list     # List(1,...)
  constant BigInt
  constant Double
  constant Boolean
  constant String
\end{lstlisting}
\caption{The \mmt Theory \cn{Scala}}\label{fig:metathy-scala}
\vspace*{-2em}
\end{wrapfigure}
Thus, we use the \mmt theory \cn{Scala} in Figure~\ref{fig:metathy-scala}, which gives mainly the important type operators and their introductory forms.
Where applicable, we use \mmt notations that mimic Scala's concrete syntax.
This has the added benefit that the resulting theory is hardly Scala-specific and thus can be reused easily for other programming languages.
It would be straightforward to add typing rules to this theory by using a logical framework as the meta-theory of \cn{Scala}, but this is not essential here.

\paragraph{Representing Classes}
It is now straightforward to represent a Scala trait $T$ containing
only 
\begin{inparaenum}[1.]
\item type members,
\item value members whose types only use symbols from \cn{Scala},
\item method members whose argument and return types only use symbols from
  \cn{Scala}
\end{inparaenum}
as an \mmt theory $\enc{T}$ with meta-theory \cn{Scala}.
\begin{compactenum}
 \item \lstinline|type n| yields \lstinline|constant n: type|
 \item \lstinline|val n: $A$| yields \lstinline|constant n: $\enctt{A}$|
 \item \lstinline|def n(x1:$A_1$,..,x_r:$A_r$):$A$| yields \lstinline|constant n: ($\enctt{A_1}$,...,$\enctt{A_r}$)=>$\enctt{A}$|
\end{compactenum}
Here $\enc{A}$ is the structural translation of the Scala type $A$ into an \mmt expression, which replaces every Scala primitive with the corresponding symbol in \cn{Scala}.

Similarly, we represent every object $O$ defining (exactly) the members of $T$ as an \mmt view $\enc{O}:\enc{T}\to\cn{Scala}$.
The member definitions in $O$ give rise to assignments in \enc{O} as follows:
\begin{compactenum}
 \item \lstinline|type n = $t$| yields \lstinline|constant n = $\enctt{t}$|
 \item \lstinline|val n: $A$ = $a$| yields \lstinline|constant n = "$a$"|
 \item \lstinline|def n($x_1$:$A_1$,...,$x_r$:$A_r$):$A$ = $a$| yields \lstinline|constant n = ($x_1$:$\enctt{A_1}$,...,$x_r$:$\enctt{A_r}$):$\enctt{A}$ = "$a$"|
\end{compactenum}

Here \lstinline|"E"| represents the escaped representation of the literal Scala
expression \lstinline|E|.  Note that we do not escape the $\lambda$-abstraction in the implementation of
\lstinline|comp|.  The resulting partially escaped term is partially parsed and analyzed by \mmt.
This has the advantage that the back-translation from \mmt to Scala can reuse the same variable names that the Scala programmer had chosen.

\begin{example}
  A Scala class for monoids (with universe, unit, and composition) and an implementation in terms of the integers are given as
  the top two code fragments in Figure~\ref{fig:monoid-scala-mmt}, their \mmt
  representations in the lower two.
\end{example}

\begin{figure}[ht]\centering
\begin{tabular}{|p{4.5cm}|p{6.8cm}|}\hline
\begin{lstlisting}
trait Monoid {
  type U
  val  unit: U
  def  comp(x1: U, x2: U): U
}
\end{lstlisting}
&
\begin{lstlisting}
object Integers extends Monoid {
  type U    = BigInt
  val  unit = 0
  def  comp(x1: U, x2: U) = x1 + x2
}
\end{lstlisting}
\\\hline
\begin{lstlisting}
theory Monoid : Scala
  constant U    : type
  constant unit : U
  constant comp : (U,U) => U
\end{lstlisting}
  &
\begin{lstlisting}
view Integers : Monoid -> Scala
  constant U    = BigInt
  constant unit = "0"
  constant comp = (x1:U, x2:U) => "x1 + x2"
\end{lstlisting}
\\\hline
\end{tabular}
\caption{Scala and \mmt representations  of Monoids and Integers}\label{fig:monoid-scala-mmt}
\end{figure}

\paragraph{Representing the Module Systems}
The correspondence between \mmt theory inclusions and Scala class extensions is not exact due to what we call the import name clash in \cite{RK:mmt:10}:
It arises when modules $M_1$ and $M_2$ both declare a symbol $c$ and $M$ imports both $M_1$ and $M_2$.
\openmath and \mmt use qualified names for scoped declarations (e.g., $M_1?c$ and $M_2?c$) so that the duplicate use of $c$ is inconsequential.
But Scala -- typical for programming languages -- identifies the two constants if they have the same type.

There are a few ways to work around this problem, and the least awkward of them is to qualify all field names when exporting \mmt theories to Scala.
Therefore, the first declaration in the trait \cn{Monoid} is actually \cn{type\;Monoid\_U} and similar for all other declarations.
Vice versa, when importing Scala classes, we assume that all names are qualified in this way.

It remains future work to align larger fragments of the module systems, which would also
include named imports and sharing.

%% file: mmt-biform.tex
We can now combine the representations of \openmath and Scala in \mmt into a biform theory graph.
In fact, we will obtain this combination as an example of a general principle of combining a logic and a programming language.

\paragraph{Bifoundations}
Consider a logic represented as an \mmt theory $L$ and a programming language represented (possibly partially as in our case with Scala) as an \mmt theory $P$.
Moreover, consider an \mmt theory morphism $s:L\to P$.
Intuitively, $s$ describes the meaning of $L$-specifications in terms of $P$.

\begin{definition}\rm
  A \emph{bifoundation} is a triple $(L,P,s:L\to P)$.
\end{definition}

\begin{wrapfigure}{r}{2cm}
\vspace{-5em}
\begin{tikzpicture}[scale=.8]
	\node (L) at (0,2) {$L$};
	\node (P) at (2,2) {$P$};
	\node (T) at (0,0) {$T$};
	\draw[-\arrowtip](L) --node[above]{$s$} (P);
	\draw[\arrowtipmono-\arrowtip](L) -- (T);
	\draw[-\arrowtip](T) --node[right]{$r$} (P);
\end{tikzpicture}
\vspace{-4em}
\end{wrapfigure}
Now consider a logical theory $T$ represented as an \mmt theory with meta-theory $L$.
This yields the diagram in the category of \mmt theories, which is given on the right.
Then, inspired by~\cite{rabe:combining:10}, we introduce the following definition of what it means to implement $T$ in $P$:

\begin{definition}\rm
  A \emph{realization} of $T$ over a bifoundation $(L,P,s)$ is a morphism $r:T\to
  P$ such that the resulting triangle commutes.
\end{definition}

Note that in \mmt, there is a canonical pushout $s(T)$ of $T$ along $s$.
Thus, using the canonical property of the pushout, realizations $r$ are in a canonical bijection with morphisms $r':s(T)\to P$ that are the identity on $P$.

\paragraph{A Bifoundation for \openmath CDs and Scala}
We obtain a bifoundation by giving an \mmt morphism $s:\cn{OpenMath}\to\cn{Scala}$.  This
morphism hinges upon the choice for the Scala type that interprets the universal type
\cn{Object}.  There are two canonical choices for this type, and the resulting morphisms
are given in Figure~\ref{fig:semsyn}. Firstly, we can choose the universal Scala type
\cn{Any}.  This leads to a semantic bifoundation where we interpret every \openmath object
by its Scala counterpart, i.e., integers as integers, lists as lists, etc. Secondly, we
can choose a syntactic bifoundation where every object is interpreted as itself.  This
requires using a conservative extension \cn{ScalaOM} of Scala that defines inductive types
\cn{Term} of \openmath objects and \cn{Context} of \openmath contexts.  Such an extension
is readily available because it is part of the \mmt API.

\begin{figure}[ht]\centering
\begin{tabular}{|p{6cm}|p{6.1cm}|}
\hline
\begin{lstlisting}
view Semantic: OpenMath -> ScalaOM
  constant Object = Any
  constant mapsto = Function
  naryObject      = List[Any]
  binder          = (Context,Term) => Any
  FMP             = (x:Any) => "assert(x == true)"
\end{lstlisting}
&
\begin{lstlisting}
view Syntactic: OpenMath -> ScalaOM
  constant Object = Term
  constant mapsto = Function
  naryObject      = List[Term]
  binder          = (Context,Term) => Term
  FMP             = (x:Term) =>
          "assert(x == OMS(logic1.true))"
\end{lstlisting}
\\\hline
\end{tabular}
\caption{Two Bifoundations From Scala to \openmath}\label{fig:semsyn}
\end{figure}

In both cases, $n$-ary arguments are easily interpreted in terms of lists and functions as functions.
The case for binders is subtle: In both cases, we must interpret binders as Scala functions that take a syntactic object in context.
Therefore, even the semantic foundation requires \cn{ScalaOM} as the codomain.

Finally, we map mathematical properties to certain Scala function calls, e.g., assertions.
In the semantic case, we assert the formula to be \cn{true}.
In the syntactic case, we assert it to be equal to the symbol \cn{true} from the \openmath CD \cn{logic1}.
Here, \cn{OMS} is part of the \mmt API.

Of course, in practice, only the simplest of FMPs actually hold in the sense that a simple Scala computation could prove them.
However, our interpretation of FMPs is still valuable: It naturally translates examples given in the \openmath CDs to Scala test cases that can be run systematically and automatically.
Moreover, in the syntactic case, we have the additional option to collect the asserted formulas and to maintain them as input for verification tools.




%% file: rules.tex
We are particularly interested in the syntactic bifoundation given above.
It corresponds to the well-understood notion of a syntactic model of a logic.
Thus, it has the advantage of completeness in the sense that the algorithms given in $T$-realizations can be used to describe deductive statements about $T$.
In this section, we make this more precise and generalize it to arbitrary logics.


\paragraph{Abstract Rewrite Rules}
First we introduce an abstract definition of rule that serves as the interface between the computational and the deductive realm.
We need one auxiliary definition:

\begin{definition}\label{def:arity}\rm
An \defemph{arity} is an element of $\{n,n\ast: n\in N\}\cup\{\cn{binder}\}$.
\end{definition}

We use $n$ ($n\ast$) for symbols that can be applied to $n$ arguments (and a sequence argument), and we use $\cn{binder}$ for symbols that form binding object.
For example, $2$ is the arity of binary symbols and $0\ast$ the arity of symbols with an arbitrary sequence of arguments.
This is a simplification of the arities we give in \cite{KR:omsemantics:12} and use in \mmt, which permit sequences anywhere in the argument list and gives binders different arities as well.

Now let us fix an arbitrary set of \mmt theories and write $\CONST$ for the set of constants declared in them.
We write $\TERM$ for the set of closed \mmt terms using only constants from $\CONST$, and $\TERM(x_1,\ldots,x_n)$ for the set of terms that may additionally use the variables $x_1,\ldots,x_n$.
Then we define:

\begin{definition}\rm
  A \defemph{rule} $r$ for a constant $c$ with arity $n\in\N$ is a mapping $\TERM^n\to\TERM$.
  Such a rule is \defemph{applicable} to any $t\in\TERM$ of the form
  $\oma{c,t_1,\ldots,t_n}$. In that case, its intended meaning is the formula $t=r(t_1,\ldots,t_n)$.

  A \defemph{rule} for a constant $c$ with arity $n\ast$ is a mapping $\TERM^n\times(\bigcup_{i=0}^\infty
  \TERM^i)\to\TERM$.  Such a rule is \defemph{applicable} to any $t\in\TERM$ of the form
  $\oma{c,t_1,\ldots,t_k}$ for $k\geq n$. In that case, its intended meaning is the formula $t=r(t_1,\ldots,t_k)$.

  A \defemph{rule} for a constant $c$ with arity $\cn{binder}$ is a mapping
  $\{(G,t)|G=x_1,\ldots,x_n\wedge t\in \TERM(G)\}\to\TERM$.
  Such a rule is \defemph{applicable} to any $t\in\TERM$ of the form $\ombind{c}{G}{t'}$.
  In that case, its intended meaning is the formula $t=r(G,t')$.

  A \defemph{rule base} $R$ is a set of rules for some constants in $\CONST$.  We write
  $R(c,a)$ for the set of rules in $R$ for the constant $c$ with arity $a$.
\end{definition}

Our rules are different from typical rewrite rules~\cite{BadNip:traat99} of the form
$t_1\rewrites t_2$ in two ways.  Firstly, the left hand side is more limited: A rule for
$c$ is applicable exactly to the terms $t_1$ whose head is $c$.  This corresponds to the
intuition of a rule implementing the constant $c$.  It also makes it easy to find the
applicable rules within a large rule base.  Secondly, the right hand side is not limited
at all: Instead of a term $t_2$, we use an arbitrary function that returns $t_2$.  This
corresponds to our open-world assumption: Constants are implemented by arbitrary programs
(written in any programming language) provided by arbitrary sources.

In the special case without binding, our rules are essentially the same as those used in
\cite{Farmer:btc07}, where the word \emph{transformer} is used for the function $r(-)$.

It is now routine to obtain a rewrite system from a rule base:

\begin{definition}\rm\label{def:rewrite}
Given a rule base $R$, $R$-rewriting is the reflexive-transitive closure of the relation $\rewrites\sq \TERM\times\TERM$ given by:

\[\mathll[c]{
\rul{r\in R(c,0)}{c \rewrites r()} \tb\tb
\rul{t_i\rewrites t_i' \mfor i=0,\ldots,n}{\oma{t_0,\ldots,t_n}\rewrites\oma{t'_0,\ldots,t'_n}} \tb\tb
\rul{r\in R(c,n) \mor r\in R(c,i\ast)\mfor i\leq n}{\oma{c,t_1,\ldots,t_n}\rewrites r(t_1,\ldots,t_n)} \\[0.3cm]
\rul{t_i\rewrites t'_i \mfor i=1,2}{\ombind{t_1}{G}{t_2}\rewrites \ombind{t'_1}{G}{t'_2}} \tb\tb
\rul{r\in R(c,\cn{binder})}{\ombind{c}{G}{t}\rewrites r(G,t)}
}\]
\end{definition}

$R$-rewriting is not guaranteed to be confluent or terminating.  This is unavoidable due
to our abstract definition of rules where not only the set of constants and rules are
unrestricted but even the choice of programming language.  However, this is usually no
problem in practice if each rule has evaluative flavor, i.e., if it transforms a more
complex term into a simpler one.

\paragraph{Realizations as Rewriting Rules}
Consider a realization $r$ of $T$ over the bifoundation
$(\cn{OpenMath},\cn{ScalaOM},\cn{Syntactic})$, and let $\rho$ be the corresponding Scala
object.  Then for every constant $c$ with type
$\oma{\cn{mapsto},\cn{Object},\ldots,\cn{Object}}$ declared in $T$, we obtain a rule $r_c$
by putting $r_c(t_1,\ldots,t_n)$ to be the result of evaluating the Scala expression
$\rho.c(t_1,\ldots,t_n)$\footnote{Technically, in practice, we need to catch exceptions and set a
  time-out to make $r_c$ a total function, but that is straightforward.}.  We obtain rules
for constants with other types accordingly. More generally, we define:
\begin{definition}\rm
  Given a theory $T$, an \defemph{arity assignment} maps every $T$-constant to an arity.
  
  Given an arity assignment, a realization $T\to \cn{ScalaOM}$ is called \defemph{syntactic} if the type of every
  $T$-constant with arity $a$ is mapped to the following Scala type:
  \texttt{(Term,\ldots,Term) => Term} if $a=n$; \texttt{(Term,\ldots,Term, List[Term]) => Term} if $a=n\ast$; and
  \texttt{(Context,Term) => Term} if $a=\cn{binder}$.
  
\end{definition}

A syntactic realization $r:T\to\cn{ScalaOM}$ induces for every constant $c$ of $T$ a rule $r_c$ in a straightforward way.
If $c$ has arity $n$, the rule $r_c$ maps $(t_1,\ldots,t_n)$ to the result of evaluating the Scala expression $r(c)(t_1,\ldots,t_n)$, where $r(c)$ is the Scala function that $r$ assigns to $c$.
Technically, $r_c$ is only a partial function because evaluation might fail or not terminate; in that case, we put $r_c(t_1,\ldots,t_n)=\oma{c,t_1,\ldots,t_n}$.
For other arities, $r_c$ is defined accordingly.

\begin{definition}\label{def:inducedbase}
We write $\mathrm{Rules}(r)$ for the rule base containing for each constant $c$ declared in $T$ the rule $r_c$.
\end{definition}

A general way of obtaining arity assignments for all theories $T$ with a fixed meta-theory $L$ is
to give an \mmt morphism $e:L\to\cn{OpenMath}$.  $e$ can be understood as a
\emph{type-erasure translation} that forgets all type information and merges all types
into one universal type.  Then the arities of the $T$-constants are determined by the
\openmath types in the pushout $e(T)$.  Therefore, we can often give bifoundations for
which all realizations are guaranteed to be syntactic, the bifoundation
$(\cn{OpenMath},\cn{ScalaOM},\cn{Syntactic})$ being the trivial example.

Def.~\ref{def:inducedbase} applies only to realizations in terms of Scala.
However, it is straightforward to extend it to arbitrary programming languages.
Of course, \mmt\ -- being written in Scala -- can directly execute Scala-based realizations whereas for any other codomain it needs a plugin that supplies an interpreter.


\paragraph{The Universal Machine}
We use the name \emph{universal machine} for the new \mmt component that maintains the rule base arising as the union of all sets $\mathit{Rules}(r)$ for all syntactic realizations $r$ with domain $\cn{ScalaOM}$ in \mmt's knowledge base.
Here ``universal'' refers to the open-world perspective that permits the extension with new logics and theories as well as programming languages and implementations.

The universal machine implements the rewrite system from Def.~\ref{def:rewrite} by exhaustively applying rules (which are assumed to be confluent) and exposes it as a single API function, called \emph{simplification}.
The \mmt system does not perform simplification at any specific point.

Instead, it is left to other components like plugins and applications to decide if and when simplification should be performed.
In the \mmt API, any term may carry metadata, and this is used to mark each subterm that has already been simplified.
Thus, different components may call simplification independently without causing multiple traversals of the same subterm.

Additionally, the API function is exposed in two ways.
Firstly, \mmt accepts simplification requests via HTTP post, where input and output are given as strings using \mmt notations or as \openmath XML elements.
Secondly, simplification is integrated with the Scala interactive interpreter, where users can type objects using \mmt notations and simplification is performed automatically.
It is straightforward to connect further frontends.

%
%


%% file: library.tex

We evaluate the new \mmt concepts by building a biform \mmt theory graph based on the bifoundation $(\cn{OpenMath},\cn{Scala},\cn{Syntactic})$, which represents $>30$ of the official OpenMath CDs in \mmt and provides Scala implementations and test cases for $>80$ symbols.
This development is available as an \mmt project and described in more detail at \url{https://tntbase.mathweb.org/repos/oaff/openmath}.

\mmt projects \cite{HIJKR:dimensions:11} already support different dimensions of knowledge, such as source, content, and presentation, as well as build processes that transform developments between dimensions.
We add one new dimension for generated programs and workflows for generating it.

Firstly, we write \mmt theories representing the \openmath CDs such as the one given on
the left of Fig.~\ref{fig:omex}.  Specifically, we represent the \cn{arith}, \cn{complex},
\cn{fns}, \cn{integer}, \cn{interval}, \cn{linalg}, \cn{list}, \cn{logic}, \cn{minmax},
\cn{nums}, \cn{relation}, \cn{rounding}, \cn{set}, \cn{setname}, and \cn{units} CDs along
with appropriate notations.

Secondly, we write views from these CDs to \cn{ScalaOM}.
Then a new \mmt build process generates all corresponding Scala classes.
Typically, users write view stubs in \mmt and then fill out the generated Scala stubs using an IDE of their choice.
Afterwards \mmt imports the Scala stubs and merges the user's changes into the \mmt views.
This is exemplified in Fig.~\ref{fig:realizations}.
Here the left side gives a fragment of an \mmt view out of \cn{arith1}, which implements arithmetic on numbers.
(We also give other views out of \cn{arith1}, e.g., for operations on matrices.)
The implementation for \cn{plus} is still missing whereas the one for \cn{minus} is present.
The right side shows the generated Scala code with the editable parts marked by comments.

\begin{tabularfigure}{|l|l|}{Partial Realization in \mmt and Generated Scala Code}{fig:realizations}
\hline
\begin{lstlisting}
view NumberArith :
  arith1 -> ScalaOM =
    plus = (args: List[Term]) "
    "
    
    minus = (a: Term, b: Term) "
      (a,b) match {
        case (OMI(x), OMI(y)) =>
          OMI(x - y)
      }
    "
\end{lstlisting}
&
\begin{lstlisting}
object NumberArith extends arith1 {
  def arith1_plus(args: List[Term]) : Term = {
    // start NumberArith?plus
    // end NumberArith?plus
  }
  def arith1_minus(a: Term, b: Term) : Term = {
    // start NumberArith?minus
    (a,b) match {
        case (OMI(x), OMI(y)) => OMI(x - y)
      }
    // end NumberArith?minus
  }
}
\end{lstlisting}
\\
\hline
\end{tabularfigure}

Finally, we write \mmt theories for extensions of the \openmath CDs with examples as on the right in Fig.~\ref{fig:omex}.
We also give realizations for them, which import the realizations of the extended CDs.
Here \mmt generates assertions for each FMP.

To apply these workflows to large libraries, we have added three build processes to \mmt that can be integrated easily with make files or \mmt IDEs.
\texttt{extract} walks over an \mmt project and translates realizations into Scala source files containing the corresponding objects. This permits editing realizations using Scala IDEs.
\texttt{integrate} walks over the Scala source files and merges all changes made to the realizations back into the \mmt files.
\texttt{load} walks over the Scala source files, compiles them, loads the class files, and registers the rule bases $\mathit{Rules}(r)$ with the universal machine. Optionally, it runs all test cases and generates a report.



%% file: conc.tex
We described a formal framework and a practical infrastructure for biform theory development, i.e., the integration of deductive theories and computational definitions of the functions specified in them.
The integration is generic and permits arbitrary logics and programming languages; moreover, the same module system is used for specifications and implementations.

We have instantiated our design with a biform development of the \openmath content dictionaries in Scala.
Future work will focus on the development of larger biform libraries and the use of further logics and programming languages.
In particular, we want to explore how to treat richer type systems and to preserve their information in the generated Scala code.

Regarding the integration of deduction and computation we focused only on ``soft verification'', i.e., linking function symbols with unverified implementations.
We only extracted the computational content of examples (which results in test cases) and omitted the more difficult problem of axioms.
We believe that future work can extend our approach to generate computation rules by spotting axioms of certain shapes such as those in inductive definitions or rewrite rules.
Moreover, given a verifier for the used programming language, it will be possible to generate the verification obligations along with the generated programs.



%% file: main-final.bbl
\newcommand{\etalchar}[1]{$^{#1}$}
\providecommand\seen{seen } \providecommand\webpageat{web page at }
  \providecommand\homepageat{home page at }
  \providecommand\projectpageat{project page at }
  \providecommand\systempageat{system home page at }
  \providecommand\svnrepoat{Subversion repository at }
  \providecommand\January{January} \providecommand\February{February}
  \providecommand\Feb{February} \providecommand\March{March}
  \providecommand\April{April} \providecommand\May{May}
  \providecommand\June{June} \providecommand\July{July}
  \providecommand\August{August} \providecommand\September{September}
  \providecommand\October{October} \providecommand\November{November}
  \providecommand\December{December} \providecommand\AUSTRALIA{Australia}
  \providecommand\ROMANIA{Romania} \providecommand\MEXICO{Mexico}
  \providecommand\ITALY{Italy} \providecommand\USA{USA}
  \providecommand\IRELAND{Ireland} \providecommand\HUNGARY{Hungary}
  \providecommand\JAPAN{Japan} \providecommand\CANADA{Canada}
  \providecommand\SPAIN{Spain} \providecommand\NETHERLANDS{Netherlands}
  \providecommand\UK{UK} \providecommand\SWEDEN{Sweden}
  \providecommand\GERMANY{Germany} \providecommand\openmath{OpenMath}
  \providecommand\fc{forthcoming} \providecommand\PROC{Proceedings}
  \providecommand\omdoc{OMDoc} \providecommand\activemath{ActiveMath}
  \hyphenation{Wiki-Sym}